# A Fully Relativistic Density Functional Study of the Actinide Nitrides

Raymond Atta-Fynn and Asok K. Ray*

*Physics Department, University of Texas at Arlington, Arlington, Texas 76019*

*akr@uta.edu.

**Abstract**

The full potential all electron linearized augmented plane wave plus local orbitals (FP- LAPW + lo) method, as implemented in the suite of software WIEN2K, has been used to systematically investigate the structural, electronic, and magnetic properties of the actinide compounds AnN (An = Ac, Th, Pa, U, Np, Pu, Am). The theoretical formalism used is the generalized gradient approximation to density functional theory (GGA-DFT) with the Perdew-Burke-Ernzerhof (PBE) exchange-correlation functional. Each compound has been studied at six levels of theory: non-magnetic (NM), non-magnetic with spin-orbit coupling (NM+SOC), ferromagnetic (FM), ferromagnetic with spin-orbit coupling (FM+SOC), anti-ferromagnetic (AFM), and anti-ferromagnetic with spin-orbit coupling (AFM+SOC). The structural parameters, bulk moduli, densities of states, and charge distributions have been computed and compared to available experimental data and other theoretical calculations published in the literature. The total energy calculations indicate that the lowest energy structures of AcN, ThN, and PaN are degenerate at the NM+SOC, FM+SOC, and AFM+SOC levels of theory with vanishing total magnetic moments in the FM+SOC and AFM+SOC cases, making the ground states essentially non-magnetic with spin-orbit interaction. The ground states of UN, NpN, PuN, and AmN are found to be FM+SOC at the level of theory used in the present computations. The nature of the interactions between the actinide metals and nitrogen atom, and the implications on 5f electron delocalization and localization are discussed in detail.

# I. Introduction

Actinides and compounds thereof continue to be highly complex and challenging areas of research from both scientific and technological points of view [1-6]. Actinide nitrides are very promising advanced fuel materials for fast breeder reactors. They are also target materials for transmutation of plutonium and minor actinides in fast reactor cores and in accelerator driven systems. In fact, actinide nitrides are under investigation for the fuels of the future fast neutron fission reactors developed in Forum Generation IV. The high density of the nitride fuel brings out more excess neutrons and has a higher potential to transmute the long lived fission products. If one considers the breeding ratio, appropriate thermophysical properties (high thermal conductivity, high melting point, high fuel density), chemical compatability with the Na coolant, and reprocessing feasibility, actinide nitrides appear to be a compromise between oxide and metal fuels. In fact, the thermal conductivity of PuN usually is in the range of 11-13 (W/m.K), that of UN being 20-23(W/m.K) compared to the values of 3-5 (W/m.K) for $UO_2$ and $PuO_2$ in the range of 800-1600 K. Actinide mononitrides are typically brittle, refractory materials with a melting point of usually greater than 2000°C. For uranium nitride, the melting point is around 2850°C and density is around 14.32 g/cm$^3$. As mentioned before, they have a higher energy neutron spectrum, respond better to demands of actinide burning and long core life [7-8]. Also, higher thermal conductivity provides margin of fuel melting and gives negative feedback because of the Doppler reactivity in unprotected loss of flow accidents. Nitrides, in fact, provide a direct path towards the self-consistent nuclear energy system, defined as a system which satisfies four objectives: energy generation, fuel breeding, confinement of the minor actinides and radioactive fission products, and nuclear reactor safety.

As a continuation of our continued study of actinide surface chemistry and physics [9], this work is concerned with detailed fundamental *ab initio* electronic structure studies of actinide nitrides. Such studies, though rather important as implied above, are relatively scarce in the literature. In contrast to metallic fuels, actinide nitrides form an isostructural series of mononitrides (AnN) with a simple rock-salt type structure. These systems are AcN, ThN, PaN, UN, NpN, PuN, and AmN. In fact, the actinide mononitrides AnN (An = Ac, Th, Pa, U, Np, Pu, Am) fall under a large family of actinides compounds with rocksalt-type structure, which include the monocarbides AnC, monopnictides AnX (X = N, P, As, Sb, Bi), monochalcogenides AnX (X = S, Se, Te), and their solutions. The UN phase diagram actually indicates that there are three thermodynamically stable phases below 400ºC, namely UN, $\alpha - U_2N_{3+x}$, (existing in the range $UN_{1.54}$ and $UN_{1.75}$) and $UN_2$. Also, of interest are nitride systems such as $(U_{0.8}Pu_{0.2})N$, (Np, Pu)N, $Th_3N_4$, $\beta - U_2N_3$, (Pu, Zr)N, (Pu, Am)N, and (Pu, Ac, Zr)N with Ac representing one of the minor actinides Np, Am, and Cm. Zr is added as a diluent and typically all these fuels form single-phase solid solutions under normal irradiation conditions. There are outstanding questions of interest in relation to these nitrides. One question, similar to the persistent question of magnetism in Pu for example [9], relates to magnetism in actinide nitrides. Experimentally, UN undergoes anti-ferromagnetic ordering with the Neel temperature at 53K. A small ordered moment and a moderate $\gamma -$ coefficient of the low-temperature specific heat indicates an itinerant character of UN magnetism. In highly distorted thin films (low temperature deposition) UN appears to exhibit a weak Pauli paramagnetism. In this regard, real structure and magnetic properties of bulk actinide nitrides and thin films need to be carefully examined, both theoretically and experimentally. Another question relates to the character of the 5f states in actinide nitrides. One photoelectron spectroscopy study of PuN indicates that the 5f states appear in the vicinity of the Fermi level, exhibiting the same type of features as in Pu metals and the 5f states are essentially delocalized. The third question relates to corrosion of actinides by water. There are, for example, uncertainties about the stability of UN in water. Some studies suggest that UN is stable in contact with boiling water and water at 300°C while other studies suggest that UN undergoes hydrolysis by superheated steam, In fact, XPS results indicate that a freshly fractured surface of UN quickly converts to $UO_2$ on exposure to liquid water or water vapor at ambient temperature. We first comment on *some* of the published literature.

Brooks and Kelly [10] performed linear-muffin-tin-orbital (LMTO) energy-band calculations in the atomic sphere approximation for UC and UN. They found that spin-orbit coupling induced a predominant orbital magnetic moment anti-parallel to the spin moment for UN. The overall results, such as the magnetic form factor, pressure dependence of the moment, and presence of large magnetic anisotropy, indicated itinerant electron behavior, though the authors indicated that more careful analysis are needed to distinguish between localized and itinerant *5f* magnetism. Using a relativistic linear muffin-tin orbital (RLMTO) method, Brooks also studied [10] the trends in the lattice parameters of the actinide nitrides from self-consistent LMTO, RLMTO, and spin-polarized LMTO calculations and interpreted the results in terms of metallic *5f – 5f* and covalent cation *5f –* anion *2p* contributions to the calculated equations of state. Large magnetovolume effects were found for NpN - AmN. Spin-orbit splitting increased the atomic volumes of NpN - AmN and the density of states at the Fermi level was found to be basically of $5f_{5/2}$ character for the paramagnetic ground states of UN-AmN. Using a semi-empirical potential, Kurosaki *et al.* [11] performed molecular dynamics (MD) simulations of the actinide nitrides (ThN, UN, NpN, and AmN) in the 300-2800 K temperature range and in the pressure range 0.1MPa - 1.5GPa to investigate their physical and thermodynamical properties. A Morse-type potential function was added to the Busing-Ida type potential to describe the ionic interactions and the authors concluded that MD simulations can successfully describe the physical properties of actinide nitrides. Petit *et al.* [12] calculated the electronic structure of AmN using self-interaction-corrected local-spin-density (SIC-LSD) approximation. They concluded that the properties of AmN are well described by a trivalent (f⁶) electronic configuration for Am ion. In a follow-up study, Petit *et al.* [12], using the same approximation, concluded that the localized $5f^3$ configuration (with the rest of the 5f states forming a band) is the most probable ground state of PuN. However, this conclusion was questioned by Havela *et al.* [13] from their reactive sputtering studies of Pu in an Ar atmosphere with variable concentration of N. They found that the $5f$ emission dominating closer to the Fermi level displayed characteristics similar to the Pu metal and that the $5f$ states can be assumed to be essentially delocalized. In a follow-up study, Rafaja *et al.* [13] have reported the structure and magnetic properties of UN thin films prepared by reactive vapor deposition at temperatures between -200 °C and +400°C providing a large variety of microstructures and observed the evolution of the $5f$ magnetism as a function of deviation from the ideal crystalinity. At low temperatures, the long-range antiferromagnetism is suppressed and a ferromagnetic component is induced producing a cluster glass type of ordering and eventually, UN indicates a weak Pauli paramagnetism. Marutzky *et al.* [14] have reported optical measurements from 1 to 10eV and magneto-optical measurements from 1 to 5 eV on an UN single crystal. Compared to the results for uranium monopnictides, they found an increased hybridization of the U (*5f*) states with the U (*6d*) and N (*2p*) states. Sheng [15] applied the linear free energy correlation model of Sverjensky and Molling [16] that correlates the standard free formation energy with thermodynamics of the corresponding metal cations to the actinide mononitrides, with the actinides treated as trivalent cations. The calculated free energies of formation and experimental data for some of the nitrides were found to be in fairly good agreement. Recently, Sedmidubsky *et al*. [8] calculated the enthalpies of formation of the actinide mononitride series using a full-potential linear augmented plane wave plus local basis (FP – LAPW + lo) [17] with the generalized gradient approximation [18] within the frame work of density functional theory (DFT). They observed a linear decrease in the enthalpies of formations from AcN to AmN, which was attributed to the stabilizing effect of the Madelung term as the bonding becomes more ionic. Except for AcN and PaN for which lattice parameters were optimized, the basic structural parameters of all the other nitrides were taken from a database [19] and ferromagnetic spin-polarized calculations at the scalar relativistic level were performed. The authors obtained good agreement with experimental data for UN and NpN enthalpies of formation but not for PuN and ThN. The authors speculated that the increasing role of electron correlations might have played a role for PuN. More recently, Shein *et al*. [20] have studied the electronic structure properties of cubic ThC, ThN, and ThO using the FP – LAPW + lo method within the GGA – DFT approximation. For perfectly stoichiometric ThN, they found good agreement with experimental values for the lattice constant, bulk modulus, and specific heat coefficient. They also noted that the bonding behavior of the ThX (X = C, N, and O) phases is a linear combination of covalent, ionic, and metallic characters. The LDA+U+SOC method has been used by Shorikov *et al.* [21] to investigate the electronic structures and magnetic state of the α – and δ – phases of metallic Pu and its compounds. For PuN, an $f^5$ configuration with a sizable magnetic moment was found. Ghosh *et al.* [22] have studied the ground state and optical properties of the americium monopnictides, AmX (X = N, P, As, Sb, and Bi), using the local density approximation LDA and LDA+U methods. They found that LDA predicted pseudogap-like behavior in AmN but LDA+U predicted semiconducting behavior with a real gap of 192 meV in AmN. Obviously, there are disagreements and discrepancies in the published literature about the actinide nitrides and ours, we believe, is the first attempt to treat all nitrides on an equal footing at various levels of theory

**II Computational methodology**

As mentioned in the abstract, we have carried out density functional calculations using the all-electron FP – LAPW + lo method as implemented in the all-electron WIEN2k code [17] in the GGA-DFT approximation [18] at six levels of theory namely, non-magnetic (NM), non-magnetic with spin-orbit coupling (NM+SOC), ferromagnetic (FM), ferromagnetic with spin-orbit coupling (FM+SOC), antiferromagnetic (AFM), and antiferromagnetic with spin-orbit coupling (AFM+SOC). In the WIEN2k code, core states are treated fully relativistically, while the valence states are treated at the scalar (without SOC) or fully relativistic (with SOC) level. SOC is included via a second variational step using the scalar relativistic eigenstates are basis, where all eigenstates with energies below 4.5 Ry are included, with the inclusion of $p_{1/2}$ orbitals [23] to account for the finite character of the $p_{1/2}$ wave function at the nucleus. Muffin-tin radii for the actinide atoms have been chosen as follows: 2.5 a.u. for Ac and Th, 2.4 a.u. for Pa, U, Np, Pu, and Am, and 1.7 a.u. for N. The parameter $R_{MT} \cdot K_{MAX} = 8$, where $R_{MT}$ is the smallest muffin radius and $K_{MAX}$ is the truncation for the modulus of reciprocal lattice vector, was used for the Fourier series expansion of the wave function in the interstitial region (this corresponds to a kinetic energy cut-off of 22.15 Ry).

To study type-I AFM configurations, consisting of alternating spin-up and spin-down ferromagnetic sheets along the [001] magnetic axis., we used an unit cell with four atoms per cell. For UN, we also used a unit cell with eight atoms per cell to study type-II AFM configuration, which consists of AFM order in the three cubic directions. In the results to follow, we will discuss the two configurations in detail. Reciprocal space integration in the first Brillouin zone is performed on a grid of 1000 k-points. Convergence of total energies with respect to the number of k-points and $K_{MAX}$ has been thoroughly checked. The total energy and the charge difference $\int (\rho_n - \rho_{n-1}) dr$ were simultaneously converged to terminate the self-consistent iterations. Convergence criteria for the energy and charge difference were 0.01 mRy and 0.0001, respectively. Energies of free atoms, which were used for the calculations of cohesive energies of the solids, were computed by placing atoms in a box of side 15 Å and only the $\Gamma$ point was used, with all other computational parameters remaining the same.

**IV Results and discussions**

In table 1, the calculated equilibrium lattice constants and bulk moduli obtained from fits to Murnaghan's equation of state [24] are presented at each of the six theoretical levels. The known experimental lattice constants of the actinide mononitrides are 9.74 a.u., 9.24 a.u., 9.25 a.u., 9.27 a.u., and 9.44 a.u. for ThN, UN, NpN, PuN, and AmN respectively [20, 25, 26]. It can be readily observed that the atomic volume contracts with the gradual filling of the 5f electrons up to U and begins to expand slowly from Np to Pu, and finally a significant increase in the atomic volume of Am. Comparing the predicted lattice constants in table 1 to the respective experimental lattice constants we find a good agreement from ThN and UN at all theoretical levels, with the maximum percent error being 0.6. Moreover, the equilibrium lattice constants of 10.45 a.u. and 9.37 a.u. obtained for AcN and PaN respectively are in excellent agreement with the recent theoretical values of 10.42 a.u. and 9.37 a.u. obtained by Sedmidubsky *et al*. [8], while the equilibrium lattice constant for of 9.79 for ThN is in exact agreement with a recent calculation by Shein *et al.* [20]. From NpN to AmN, the calculated lattice constants at the AFM, AFM+SOC, FM, FM+SOC levels of theory are also in good agreement with experiments, the maximum percent error being all less than unity. For NpN-AmN, significant discrepancies (greater than 1%) is observed at the NM (and NM+SOC) levels theory, with AmN showing the largest departure (5%) from the experimental value. This tends to indicate that unlike ThN and UN and possibly AcN and PaN, a spin-polarized theory is needed to accurately predict the lattice parameters of NpN to AmN. Our equilibrium lattice constant of 9.29 a.u. for PuN is in much better agreement with the experimental value of 9.27 a.u. compared to the value of 9.69 a.u. by obtained Petit *et al*. [12]. Also, our equilibrium lattice constant of 9.40 and a value of 9.44 a.u. obtained by Petit *et al*. [12] agree well with the experimental value of 9.44 a.u. for AmN, while the value of 9.12 a.u. obtained by Ghosh *et al.* [22] underestimates the experimental value by 3.8%. Further comparisons of lattice constants are summarized in figure 1, where we show a plot of the lattice constants obtained for our lowest energy structures, the known experimentally measured lattice constants, and that of Brooks [10]. For ThN, Brooks' results indicate a 2.1% overestimation of the experimental value whereas our result overestimate the experimental value by 0.4%. For UN, our lattice constant and Brooks' are underestimated by 0.3% and 1.2%, respectively. Brook's results for NpN is almost in exact agreement with experiment whereas ours show a 0.6% contraction. Both results for PuN show a small expansion of about 0.2% while the results for AmN show contractions 0.4 % in our case and 2.3% in Brooks' results. Overall, the lattice constants obtained in both cases agree reasonably well with experiment. Next we discuss the trends in the bulk moduli reported in table 1. For AcN-UN, the reported values are fairly consistent at each theoretical level and the respective percent errors from experimental data, available only for ThN, varies from 1.7 to 7.4 percent. Compared to Brooks' results (shown in parenthesis), the bulk moduli in GPa corresponding to our lowest energy structures are respectively: 99, 170 (174), 200 (217), 219 (214), 183 (200), 147 (194), and 145 (177) for AcN, ThN, PaN, UN, NpN, PuN, and AmN. The results for ThN, PaN, UN, and NpN are in fair agreement with each other while significant discrepancies are observed for PuN and AmN.

In the first column of table 2, the total energy difference $\Delta E$ for each compound from their respective ground states is listed for each theoretical level. Here, $\Delta E > 0$ indicates instability. First of all, it is clearly evident that with and without SOC, the ground states of AcN to PaN at the NM, AFM, FM (NM+SOC, AFM+SOC, FM+SOC) levels of theory are degenerate within the range 0.1-0.3 mRy. For each compound however, the total energy is lowered significantly with the inclusion of spin-orbit coupling. From the results for UN reported in table 2, we see an energy difference of about 1 mRy between the AFM+SOC and FM+SOC energies. For PuN to AmN, we clearly observe an FM+SOC ground state, with levels of theory in the increasing order stability being: NM, AFM, FM, NM+SOC, AFM+SOC, FM+SOC. A quick glance at the energy differences indicate that both spin-polarization and spin-orbit coupling effects have non-negligible influences on lowering the total energies of the UN, NpN, PuN, and AmN.

The energy differences stemming from spin-polarized and spin-orbit coupling effects are listed in table 2. To facilitate our discussions, we have pictorially represented the spin-polarization energy $E_{SP}$ in figure 2 and the spin-orbit coupling energy $E_{SO}$ in figure 3. Here $E_{SP}$ is defined as $E_{NM(+SOC)} - E_{FM(+SOC)/AFM(+SOC)}$ and $E_{SO}$ is defined as $E_{NM/AFM/FM} - E_{NM+SOC/AFM+SOC/FM+SOC}$. In figure 2, we clearly observe little or no spin-polarization effects on the energy from AcN to PaN. From UN to AmN, $E_{SP}$ increases smoothly. It is worth noting that spin-polarization lowers the total energy of the FM and AFM configurations more than it does for FM+SOC, AFM+SOC. In figure 3, we observe, at the three levels of theory with SOC, that the SOC energy for AcN, ThN, PaN, and UN is nearly constant. From NpN to AmN, we observe pronounced spin-orbit coupling energy-lowering effects at the NM+SOC level of theory, followed by AFM+SOC, and then FM+SOC.

As stated earlier, for UN, we also studied a larger cell (8 atoms/unit cell) to study type-II AFM ordering i.e., AFM ordering in all three cubic directions. This unit cell also naturally accommodates type-I AFM ordering i.e., alternating spin and down ferromagnetic sheets along [001]. We show, as an example, the result for an 8-atom UN unit cell and compare with the results obtained with the 4-atom unit cell results for the lattice constants and bulk moduli in table 3a and the energy differences in table 3b. Upon examining table 3a, we first observe that the lattice constants at the 6 theoretical levels used for the cells agree quite well, with the differences attributed to numerical inaccuracies. Also, a comparison of the AFM type-II structural properties with AFM type-I properties for the 8 atom cell indicate basically identical results. Next we look at the energy differences listed in table 3b. If we compare the energy differences at the 6 theoretical levels used for both cells, we see that the total energy differences are typically less than a mRy. The same is true for the spin-polarization energy, spin-orbit coupling energy, as well as the total and site-projected spin magnetic moments. Also, the results indicate that without SOC, the total energies of AFM type-II and AFM type-I configurations are basically degenerate, while with SOC the total energy for AFM type-I configuration is 0.838 mRy/unit cell lower than that of AFM type-II. The total energies of the ground states for the 4-atom and 8-atom cells at the FM+SOC level are -56276.131084 Ry/unit cell and -56276.133284 Ry/unit cell, respectively.

In figure 4, we show plots of the cohesive energies per unit cell $E_{cohes}$ for each solid for the lowest energy structures in comparison to recent spin-polarized GGA calculations using same code as reported in Ref. [8]. Here we define $E_{cohes}$ as

$$\frac{1}{N_U}\left[E_{An} + E_N - E_{AnN}\right],$$ where $N_U$ is the number of unit cells, $E_{An}$ and $E_N$ are respectively the isolated atomic energies of the actinide and nitrogen, and $E_{AnN}$ is the total energy of of AnN. With this definition, positive cohesive energy indicates binding and negative cohesive energy otherwise. Figure 4 indicates an increase in cohesive energies from AcN to PaN and a linear decrease from PaN to AmN while the results of Sedmidubsky *et al.*

[8] indicate a non-linear decrease from ThN to AmN. A possible explanation for the differences in the magnitudes of the cohesive energies is the difference in computational approaches used. For example, in ref. [8], only the lattice constants of AcN and PaN were optimized and also, all calculations were done using spin-polarized GGA (with SOC neglected) while in our case all lattice constants were optimized at all levels. The high cohesive energy of PaN indicate a significant contribution of the 5f and 6d to covalent bonding, with the contributions 5f electrons to covalent bonding decreasing from PaN-AmN [8].

In table 2, we have listed the net spin magnetic moments of the solids. In table 4, the site projected moment S, orbital moment L, and total moment J for each actinide with a non-zero magnetic moment corresponding to lowest energy structures (i.e. UN, NpN, PuN, and AmN at the FM+SOC level of theory) are reported. All the moments in table 4 are computed within the muffin-tin and therefore exclude interstitial contributions. Orbitals moments were computed inside the muffin tin for cases with SOC but without orbital polarization. As stated earlier, the ground states of AcN, ThN, and PaN are clearly non-magnetic. For ThN, this is in agreement with the experimentally observed paramagnetic ground state [27] and a recent theoretical calculation [20] using the same code. As mentioned before, experimental data indicates that magnetic ordering in UN occurs at the Neel temperature $T_N = 53$ K to an antiferromagnetic type-I structure [28-31]. Our results predict a ferromagnetic structure for UN at 0 K, with the total spin magnetic moment in the unit cell for the lowest energy FM+SOC structure to be $0.96\mu_B$. From table 4, we observe a cancellation of the projected spin moment by the orbital moment for U. The vanishing total moment for U is a clear indication of the delocalized nature of the U 5f electrons in UN. Next we consider the magnetic structure of NpN. Experimental studies have shown that NpN is a ferromagnet with Curie temperature $T_C$ = 87 K [32]. Our FM+SOC results reported for NpN clearly agree with experimentally observed FM ground state. From table 2 we see that the total spin magnetic moment in the cell 2.45 $\mu_B$. Again from table 4, we see a cancellation of the site projected spin moment of Np by the orbital moment, and hence just like U, the almost vanishing total moment in the muffin-tin signifies 5f delocalization in NpN. Neutron diffraction experiments for PuN showed no long-range order or magnetic moments larger than 0.25 $\mu_B$ [33]. AFM ordering at $T_N$ = 13 K was suggested on the basis of a maximum in the magnetic susceptibility and specific heat [33]. According to another magnetic susceptibility curve, PuN is a Curie-Weiss paramagnet with an effective moment $\mu_{eff}$ =1.08 $\mu_B$/Pu [34]. However, our results clearly predict a ferromagnetic ground state. From table 2, the total spin magnetic moment of the cell is 4.26 $\mu_B$. From table 4 the total Pu magnetic moment is 1.81 $\mu_B$, indicating localized 5f electron moments. The predicted magnetic moment for Pu agrees with the value of 2.06 $\mu_B$ obtained by Shorikov *et al.* [21]. For AmN, one experimental study predicts temperature independent paramagnetism [19]. However, our calculations clearly predict a ferromagnetic ground state. A total spin moment of 5.64 $\mu_B$ is reported in table 2. From table 4, we observe a small orbital magnetic moment contribution, leading a total site magnetic moment of 4.33 $\mu_B$, which is a clear sign of 5f localized moments.

We now discuss the electronic structures of the actinide compounds by focusing on the angular momentum decomposed (partial) density of states (DOS) inside the muffin tin for the f and d states for actinides and the p states of N. Only the lowest energy structures were considered, and for spin-polarized calculations, partial DOS for each angular momentum were summed over spins. In figure 5, we show the partial DOS for AcN. From about -4.0 eV up to the Fermi level, we clearly observe N 2p and Ac 6d hybridizations and very small mixing with Ac 5f states. Also, there is a very small density of the states at the Fermi level. The character and admixture of the Th 6d and N 2p states below the Fermi level for ThN shown in figure 6 is similar to the plot in figure 5 for AcN. Again, we observe a small contribution to the DOS by the 5f states and a small DOS at the Fermi level. In figure 7 we show the partial DOS for PaN. The Pa-N interaction is dominated Pa 6d, Pa 5f and N 2p hybridizations. However, we also note an appreciable density of 5f states in the in the valence region near the Fermi level. In figure 8, we depict the partial DOS for UN. Compared to Ac, Th, and Pa, we clearly observe a significant density of states at the Fermi level and significant U(5f)-N(2p), U(6d)-N(2p), and U(5f)-U(6d) hybridizations. The partial DOS for NpN, PuN, and AmN are reported in figures 9, 10, and 11 respectively. The N 2p hybridizations with the respective actinide 5f and 6d states, and weak actinide 5f-6d hybridizations are noted. The LDA predicted pseudogap-like behavior and LDA+U predicted semiconducting behavior with a real gap of 192 meV in AmN observed by Ghosh *et al.* [22] has not been observed in our DOS for AmN.

One noticeable feature in the partial DOS is the gradual increase in the density of 5f states and the appearance peaks below the Fermi level from Pa to Am, all of which have non-empty 5f orbitals. The 5f DOS for Pa is relatively small below and at the Fermi level. For U 5f partial DOS in figure 8, we see a small peak just above the Fermi level. For Np 5f partial DOS in figure 9, we clearly see a single peak below the Fermi level, while for the 5f partial DOS for Pu in figure 10, we see a peak at around -1.8 eV and one peak just below the Fermi level. For the Am 5f DOS in figure 11, we see that the peaks are withdrawn from the Fermi level. For pure Pa, U and Np 5f states, it is well known that the 5f states are delocalized and the same trend is manifested in the partial DOS for PaN, UN and NpN. Similarly, the Pu and Am 5f peaks we have observed here is similar to the behavior in their pure states, and hence, it may be interpreted as a sign of 5f electron localization.

The partial DOS was computed solely within the muffin-tins and do not tell us anything about the interaction in the interstitial region. To further elucidate the nature of the interaction between the actinide metal and N, we computed the difference charge density, which gives us information about the nature of the chemical bonds formed as result of charge redistribution. We define the difference charge density $\Delta n(r)$ as follows:

$\Delta n(r) = n(AnN) - n(An) - n(N)$,

where $n(AnN)$ is the total charge density of the AnN solid, $n(An)$ is the total charge density of the actinide metal, and $n(N)$ is the total charge density of the N atom. Furthermore, the positions of the An and N atoms remained at exactly the same positions as they were in the solid. In figure 12, we show the difference density plots for all the compounds. The density was computed in the (001) plane and the coloring scheme and scale are indicated in the figure. In general, we observe charge accumulation around N and charge depletion around the actinide atoms. This clearly suggests that there is charge transfer from the actinide to N, indicating that the chemical An-N chemical bonds are mainly ionic in character. This is expected since the N atom is more electronegative than the metal actinide atoms. Looking at the difference density plots for UN, NpN, PuN, and AmN in figure 12, charge depletion around the actinide atoms can be clearly observed.

However, this is not the case AcN, ThN, and PaN. For PaN in particular, which is the most stable in terms of cohesive energy, we see, in addition to regions of charge loss, small regions of charge gain around Pa, which we attribute to the extra contribution of covalent bonding to ionic bonding. For the other solids, we do not clearly see signs of covalent bonding. This might be due to the fact that covalent features of the bonds may be quite subtle and are not easily captured by difference density plots.

We also attempted to calculate the electronic specific heat coefficient γ by using an approximate model. For non-interacting electrons, the electronic specific heat coefficient γ is proportional to the total density of states $N(E_F)$ at the Fermi level and is given by $\gamma = \frac{\pi^2}{3} N(E_F) K_B^2$. For ThN our estimated value of 2.64 mJ / (mol $K^2$) is comparable to the experimental value of 3.12 mJ / (mol $K^2$) [27] and a recent FP-LAPW+lo with the GGA calculation [20] which yielded a value of 2.74 mJ / (mol $K^2$). But for UN and PuN, we observe a large discrepancy between our estimated values and experimental data.  To match theoretical electronic specific heat coefficient γ to experimental values, it should be corrected as  γ(corrected) = γ (band)*(1 + λ) where the parameter λ takes into account, the electron-phonon interactions and many body effects, and γ(band) is our computed value [37]. We intend to pursue such studies in the future.

## IV. Conclusions

The full potential all electron linearized augmented plane wave plus local orbitals (FP – LAPW + lo) method has been used to systematically study the zero temperature structural and electronic properties  of AnN (An = Ac, Th, Pa, U, Np, Pu, Am) at six different configurations, namely: nonmagnetic, ferromagnetic, anti-ferromagnetic, non-magnetic with spin-orbit coupling, ferromagnetic with spin-orbit coupling and anti-ferromagnetic with spin-orbit coupling. The optimized lattice constants and bulk moduli for the lowest energy structures are in good agreement with known experimental data. The ground states of AcN, ThN, and PaN are clearly non-magnetic with spin-orbit coupling. The ground states of UN, NpN, PuN, and NpN are all ferromagnetic with spin-orbit coupling. Our zero temperature ferromagnetic structures predicted for UN, PuN, and AmN contradict experimental results whereas both the ferromagnetic structure of NpN and non-magnetic structure of ThN agree with experiment. A study of the site projected magnetic moments show a cancellation of the spin and orbital moments behavior U and Np, and localized magnetic moment for Pu and Am. A study of the partial density of states showed actinide 6d and N 2p hybridizations, with some admixture from PaN, UN, PuN, and AmN 5f electrons and also a 5f electron delocalization for Pa, U, and Np and 5f localization for Pu and Am. The observed chemical bonding between the actinides and nitrogen has significant ionic character. The specific heat coefficients have been computed using the free electron model. Further experimental and theoretical work needs to be done to clarify the discrepancies with measured magnetic structures for UN, PuN, and AmN as also thermal properties such as electronic specific heat coefficients. This could imply going beyond the techniques of standard density functional theory but it is not clear at this point that this will necessarily solve the discrepancies.

**Acknowledgments**

This work is supported by the Chemical Sciences, Geosciences and Biosciences Division, Office of Basic Energy Sciences, Office of Science, U. S. Department of Energy (Grant No. DE-FG02-03ER15409) and the Welch Foundation, Houston, Texas (Grant No. Y-1525).

Table 1: Optimized lattice constants *a* (in a.u) and bulk moduli *B* (in GPa). *Δa* and *ΔB* are the respective percent error of the lattice constants and bulk moduli from experimental values.

| | Theory | *a* (a.u.) | Δ *a* (%) | *B* (GPa) | Δ *B* (%) |
|---|---|---|---|---|---|
| AcN | NM | 10.47 | | 100 | |
| | AFM | 10.47 | | 101 | |
| | FM | 10.47 | | 100 | |
| | NM+SOC | 10.45 | | 100 | |
| | AFM+SOC | 10.45 | | 99 | |
| | FM+SOC | 10.45 | | 99 | |
| ThN | NM | 9.79 | 0.5 | 178 | 1.7 |
| | AFM | 9.79 | 0.5 | 178 | 1.7 |
| | FM | 9.79 | 0.5 | 178 | 1.7 |
| | NM+SOC | 9.78 | 0.4 | 162 | -7.4 |
| | AFM+SOC | 9.78 | 0.4 | 170 | -2.9 |
| | FM+SOC | 9.78 | 0.4 | 162 | -7.4 |
| PaN | NM | 9.37 | | 223 | |
| | AFM | 9.37 | | 223 | |
| | FM | 9.37 | | 224 | |
| | NM+SOC | 9.35 | | 205 | |
| | AFM+SOC | 9.35 | | 200 | |
| | FM+SOC | 9.36 | | 210 | |
| UN | NM | 9.18 | -0.6 | 227 | |
| | AFM | 9.20 | -0.4 | 213 | |
| | FM | 9.20 | -0.4 | 209 | |
| | NM+SOC | 9.20 | -0.4 | 229 | |
| | AFM+SOC | 9.21 | -0.3 | 221 | |
| | FM+SOC | 9.21 | -0.3 | 219 | |
| NpN | NM | 9.06 | -2.1 | 228 | |
| | AFM | 9.21 | -0.4 | 177 | |
| | FM | 9.22 | -0.3 | 151 | |
| | NM+SOC | 9.13 | -1.3 | 209 | |
| | AFM+SOC | 9.19 | -0.6 | 195 | |
| | FM+SOC | 9.19 | -0.6 | 183 | |
| PuN | NM | 9.00 | -2.9 | 218 | |
| | AFM | 9.30 | 0.3 | 155 | |
| | FM | 9.36 | 1.0 | 1.48 | |
| | NM+SOC | 9.11 | -1.7 | 192 | |
| | AFM+SOC | 9.26 | -0.1 | 160 | |
| | FM+SOC | 9.29 | 0.2 | 147 | |
| AmN | NM | 8.97 | -5.0 | 211 | |
| | AFM | 9.42 | -0.2 | 130 | |
| | FM | 9.45 | 0.1 | 145 | |
| | NM+SOC | 9.12 | -3.4 | 186 | |
| | AFM+SOC | 9.36 | -0.8 | 141 | |
| | FM+SOC | 9.40 | -0.4 | 145 | |

Table 2. Total energy difference relative to the ground state ΔE (mRy/unit cell), spin-polarization energies $E_{sp}$ (mRy/unit cell), spin-orbit coupling energies $E_{so}$ (mRy/unit cell) and total spin magnetic moments $\mu_s$ ($\mu_B$/unit cell) for the actinide mononitrides at different theoretical levels.

| | Theory | $\Delta E$ (mRy/unit cell) | $E_{sp}$ (mRy/unit cell) | $E_{so}$ (mRy/unit cell) | $\mu_s$ ($\mu_B$/unit cell) |
|---|---|---|---|---|---|
| AcN | NM | 285.784 | - | - | - |
| | AFM | 285.794 | -0.010 | - | 0.00 |
| | FM | 285.791 | -0.007 | - | 0.00 |
| | NM+SOC | 0.130 | - | 285.654 | - |
| | AFM+SOC | 0.186 | -0.056 | 285.608 | 0.00 |
| | FM+SOC | 0.000 | 0.130 | 285.791 | 0.00 |
| ThN | NM | 336.245 | - | - | - |
| | AFM | 336.245 | 0.000 | - | 0.00 |
| | FM | 336.245 | 0.000 | - | 0.00 |
| | NM+SOC | 0.260 | - | 335.985 | - |
| | AFM+SOC | 0.000 | 0.260 | 336.245 | 0.00 |
| | FM+SOC | 0.002 | 0.258 | 336.243 | 0.00 |
| PaN | NM | 385.740 | - | - | - |
| | AFM | 385.740 | 0.000 | - | 0.00 |
| | FM | 385.737 | 0.003 | - | 0.00 |
| | NM+SOC | 0.223 | - | 385.517 | - |
| | AFM+SOC | 0.000 | 0.223 | 385.740 | 0.00 |
| | FM+SOC | 0.109 | 0.114 | 385.628 | 0.00 |
| UN | NM | 446.168 | - | - | - |
| | AFM | 443.706 | 2.642 | - | 0.00 |
| | FM | 440.696 | 5.472 | - | 1.23 |
| | NM+SOC | 4.433 | - | 441.735 | - |
| | AFM+SOC | 1.073 | 3.360 | 442.633 | 0.00 |
| | FM+SOC | 0.000 | 4.433 | 440.696 | 0.96 |
| NpN | NM | 535.332 | - | - | - |
| | AFM | 505.212 | 30.120 | - | 0.00 |
| | FM | 499.024 | 36.308 | - | 3.11 |
| | NM+SOC | 20.803 | - | 514.529 | - |
| | AFM+SOC | 3.569 | 17.234 | 501.643 | 0.00 |
| | FM+SOC | 0.000 | 20.803 | 499.024 | 2.45 |
| PuN | NM | 646.606 | - | - | - |
| | AFM | 553.425 | 93.181 | - | 0.00 |
| | FM | 542.775 | 103.831 | - | 4.85 |
| | NM+SOC | 37.813 | - | 608.793 | - |
| | AFM+SOC | 3.7915 | 34.022 | 549.634 | 0.00 |
| | FM+SOC | 0.000 | 37.813 | 542.775 | 4.26 |
| AmN | NM | 792.877 | - | - | - |
| | AFM | 600.031 | 192.846 | - | 0.00 |
| | FM | 583.163 | 209.714 | - | 6.00 |
| | NM+SOC | 85.125 | - | 707.752 | - |
| | AFM+SOC | 10.201 | 74.924 | 589.83 | 0.00 |
| | FM+SOC | 0.000 | 85.125 | 572.962 | 5.64 |

Table 3a: Comparison of lattice constants *a* and bulk moduli B for a 4 atom-cell and a 8-atom cell for UN. The numbers in parenthesis denote the percent error of the quantity from experimental values.

| Theory | 4 atom-cell | | 8 atom-cell | |
|---|---|---|---|---|
| | A (a.u.) | B (GPa) | Lattice constant (a.u.) | Bulk modulus (GPa) |
| NM | 9.18 (-0.6 %) | 227 | 9.20 (-0.4 %) | 229 |
| AFM2 | - | - | 9.21 (-0.3 %) | 216 |
| AFM1 | 9.20 (-0.4 %) | 213 | 9.21 (-0.3 %) | 216 |
| FM | 9.20 (-0.4 %) | 209 | 9.22 (-0.2 %) | 214 |
| NM+SOC | 9.20 (-0.4 %) | 229 | 9.20 (-0.4 % ) | 223 |
| AFM2+SOC | - | - | 9.21 (-0.3 %) | 229 |
| AFM1+SOC | 9.21 (-0.3 % ) | 221 | 9.21 (-0.3 %) | 222 |
| FM+SOC | 9.21 (-0.3 %) | 219 | 9.21 (-0.3 %) | 235 |

Table 3b: Comparisons of total energy differences relative to the ground state ΔE (mRy/unit cell), spin-polarization energies $E_{sp}$ (mRy/unit cell), spin-orbit coupling energies $E_{so}$ (mRy/unit cell) for the 4-atom and 8-atom cells.

| Theory | 4 atom-cell | | | 8 atom-cell | | |
|---|---|---|---|---|---|---|
| | $\Delta E$ (mRy/u.c) | $E_{sp}$ (mRy/u.c.) | $E_{so}$ (mRy/u.c) | $\Delta E$ (mRy/u.c.) | $E_{sp}$ (mRy/u.c.) | $E_{so}$ (mRy/u.c.) |
| NM | 446.168 | - | - | 446.145 | - | - |
| AFM2 | - | - | - | 443.501 | 2.644 | - |
| AFM1 | 443.706 | 2.642 | - | 443.503 | 2.643 | - |
| FM | 440.696 | 5.472 | - | 440.450 | 5.695 | - |
| NM+SOC | 4.433 | - | 441.7345 | 5.610 | - | 440.535 |
| AFM2+SOC | - | - | - | 2.025 | 3.585 | 441.4763 |
| AFM1+SOC | 1.073 | 3.360 | 442.633 | 1.187 | 4.424 | 442.316 |
| FM+SOC | 0.000 | 4.433 | 440.6960 | 0.000 | 5.610 | 440.450 |

Table 4: Site projected spin magnetic moment S (in $\mu_B$), orbital magnetic moment L (in $\mu_B$), and total magnetic moment (in $\mu_B$) inside the muffin-tin for the actinide atoms in lowest energy structure with magnetic ground state.

|     | S    | L     | Total |
| --- | ---- | ----- | ----- |
| U   | 0.86 | -0.85 | 0.01  |
| Np  | 2.26 | -2.21 | 0.05  |
| Pu  | 3.97 | -2.16 | 1.81  |
| Am  | 5.34 | -1.01 | 4.33  |

Table 5: Electronic specific heat coefficient $\gamma$ (in mJ mol$^{-1}$ K$^{-2}$) computed using the free electron model and the corresponding experimental values.

|                      | AcN  | ThN      | PaN  | UN        | NpN  | PuN       | AmN  |
| -------------------- | ---- | -------- | ---- | --------- | ---- | --------- | ---- |
| Computed $\gamma$    | 0.12 | 2.64     | 2.36 | 13.5      | 8.69 | 9.30      | 8.06 |
| Experimental $\gamma$ |      | 3.12 [27] |      | 25.0 [35] |      | 66.0 [36] |      |

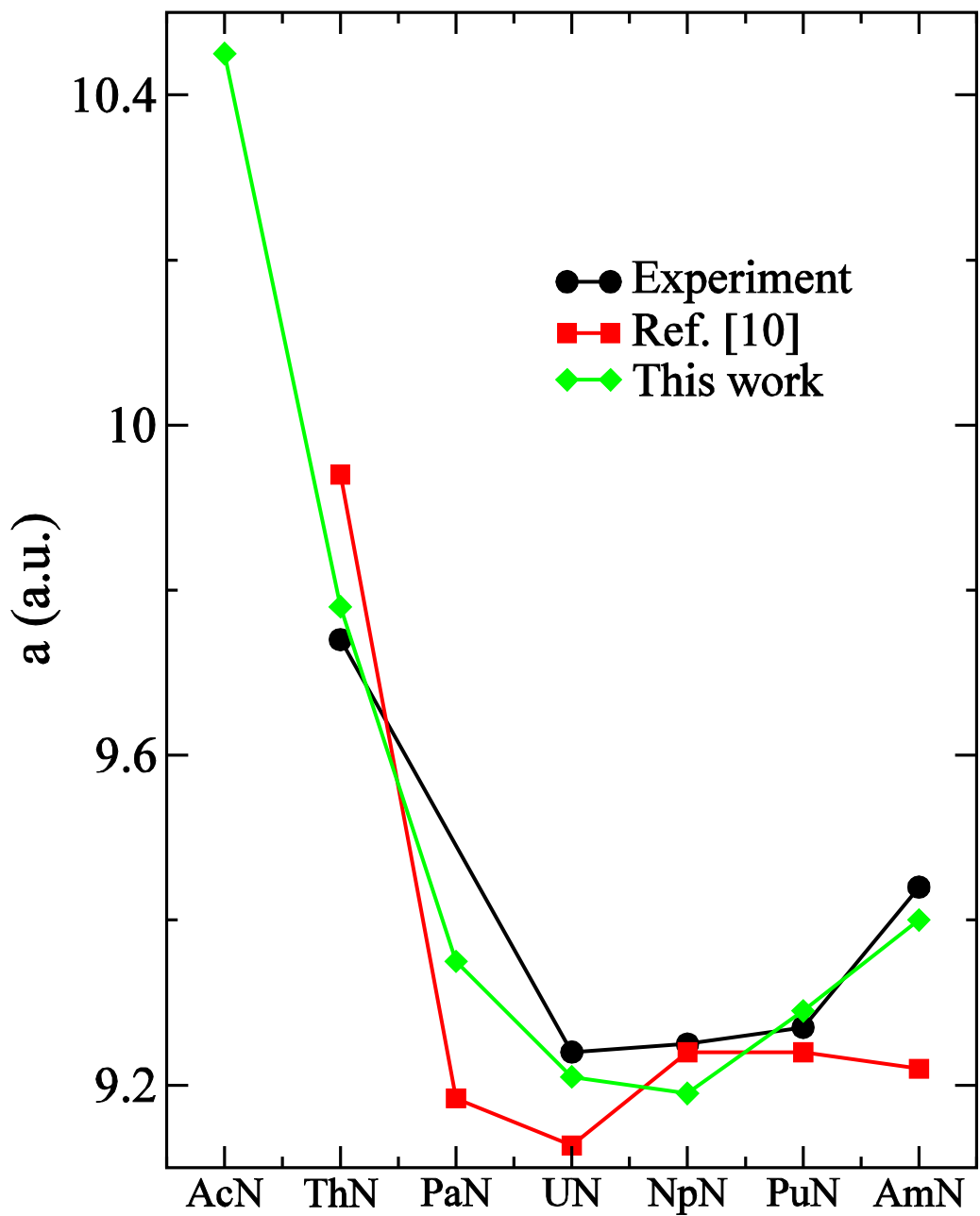

Figure 1 (Color Online): Lattice constants (in a.u.) of the actinide mononitrides.

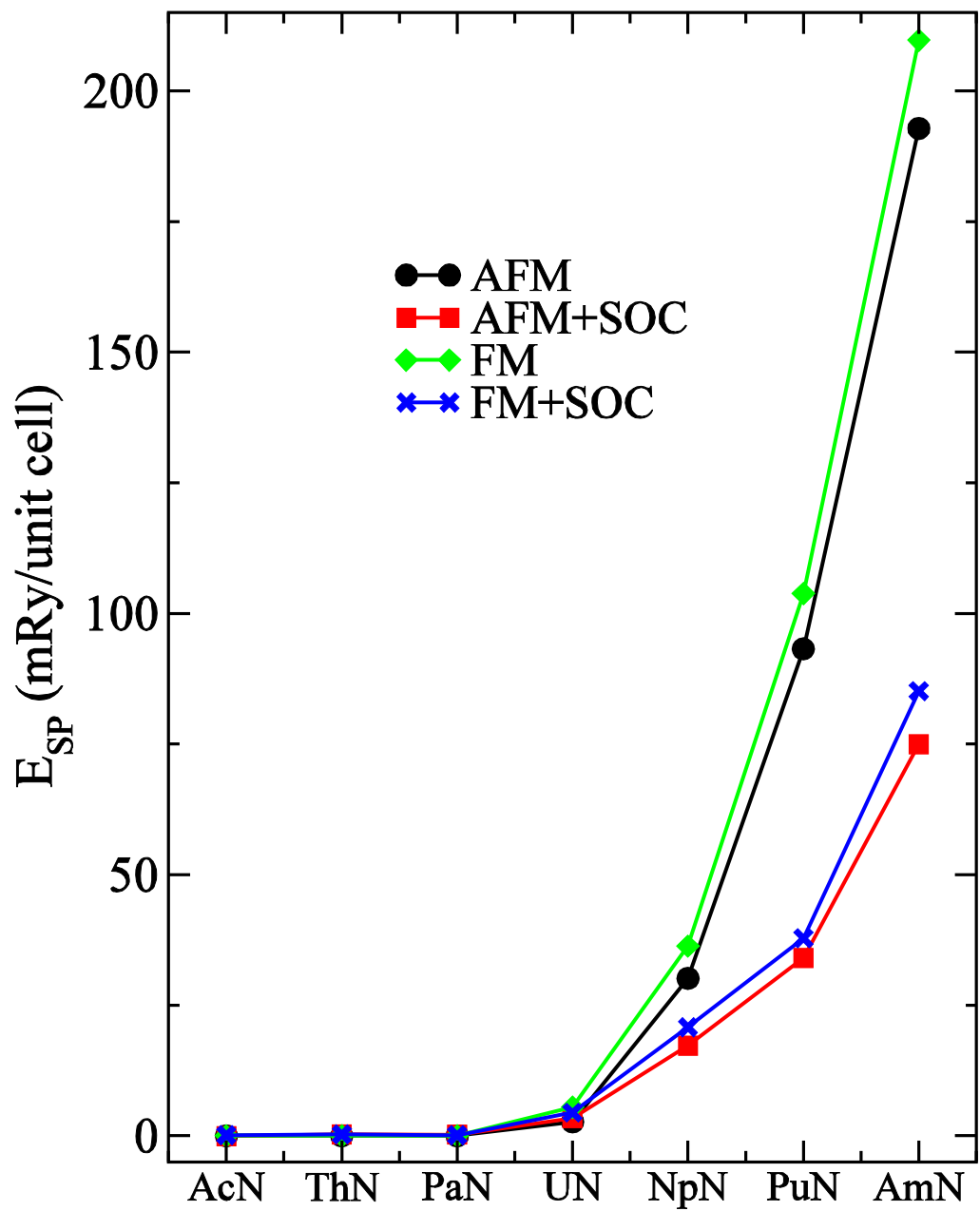

Figure 2 (Color Online): Spin-polarization energy of the actinide mononitrides.

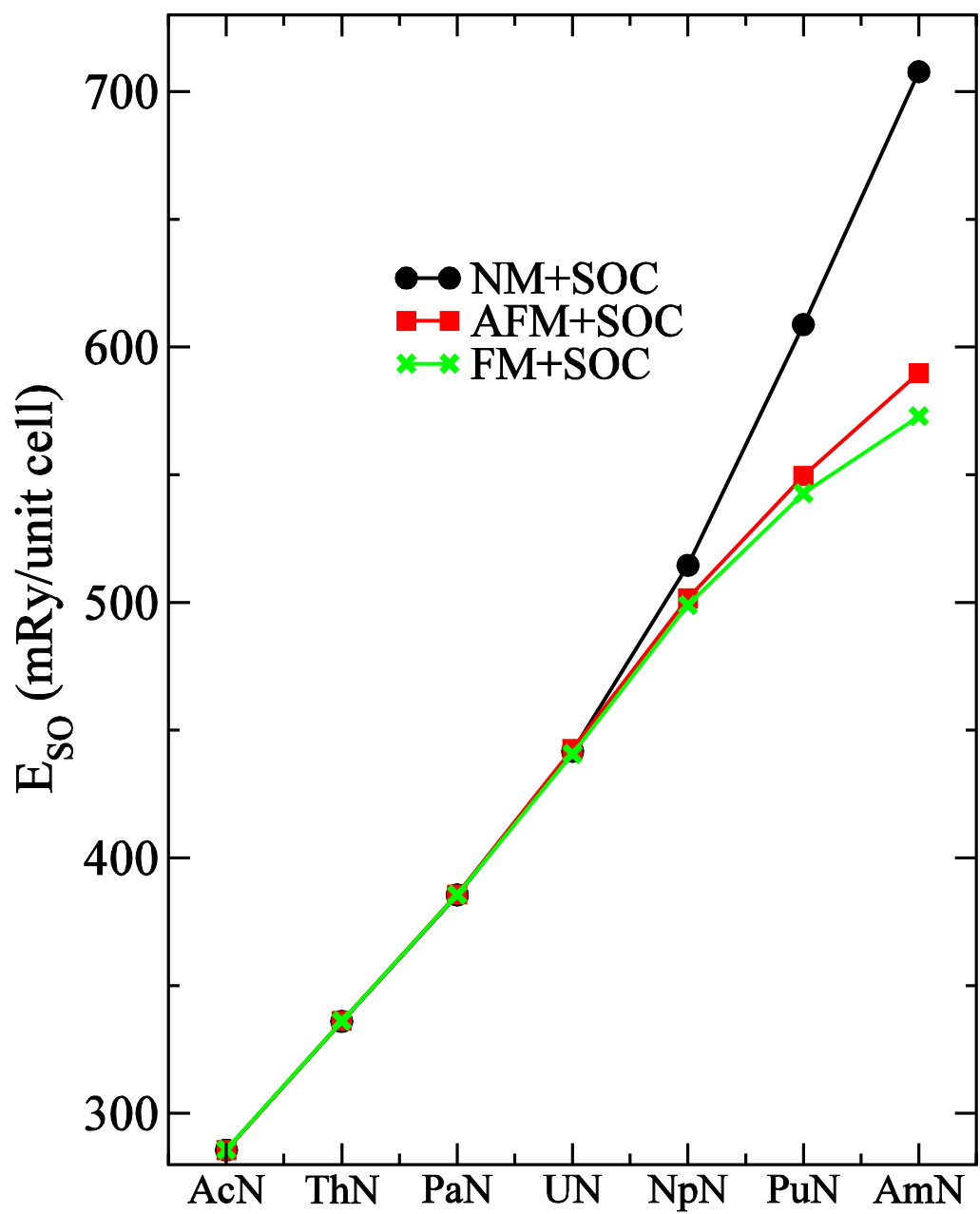

Figure 3 (Color Online): Spin-orbit coupling energy of the actinide mononitrides.

Figure 4 (Color Online): Cohesive energies of the actinide mononitrides.

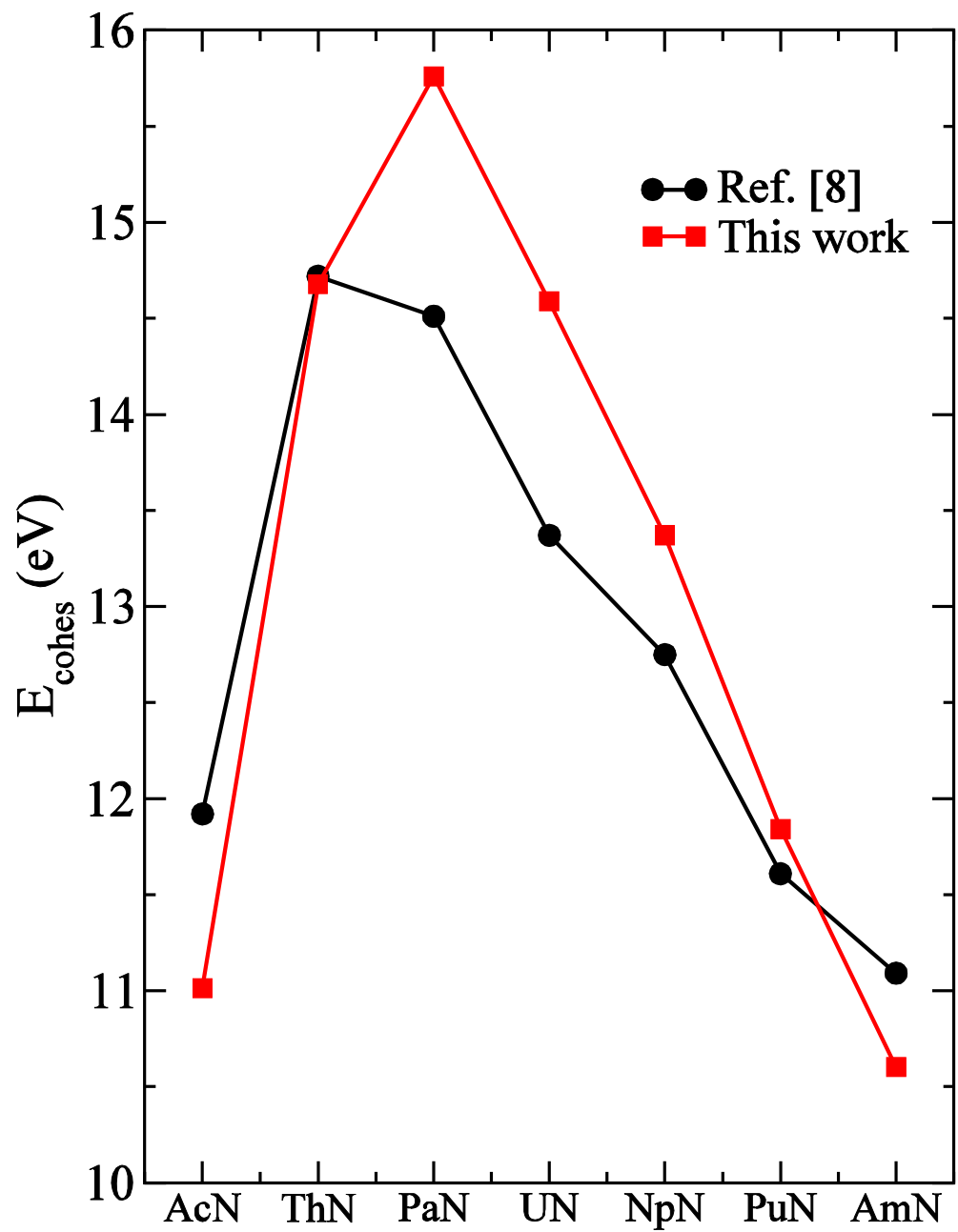

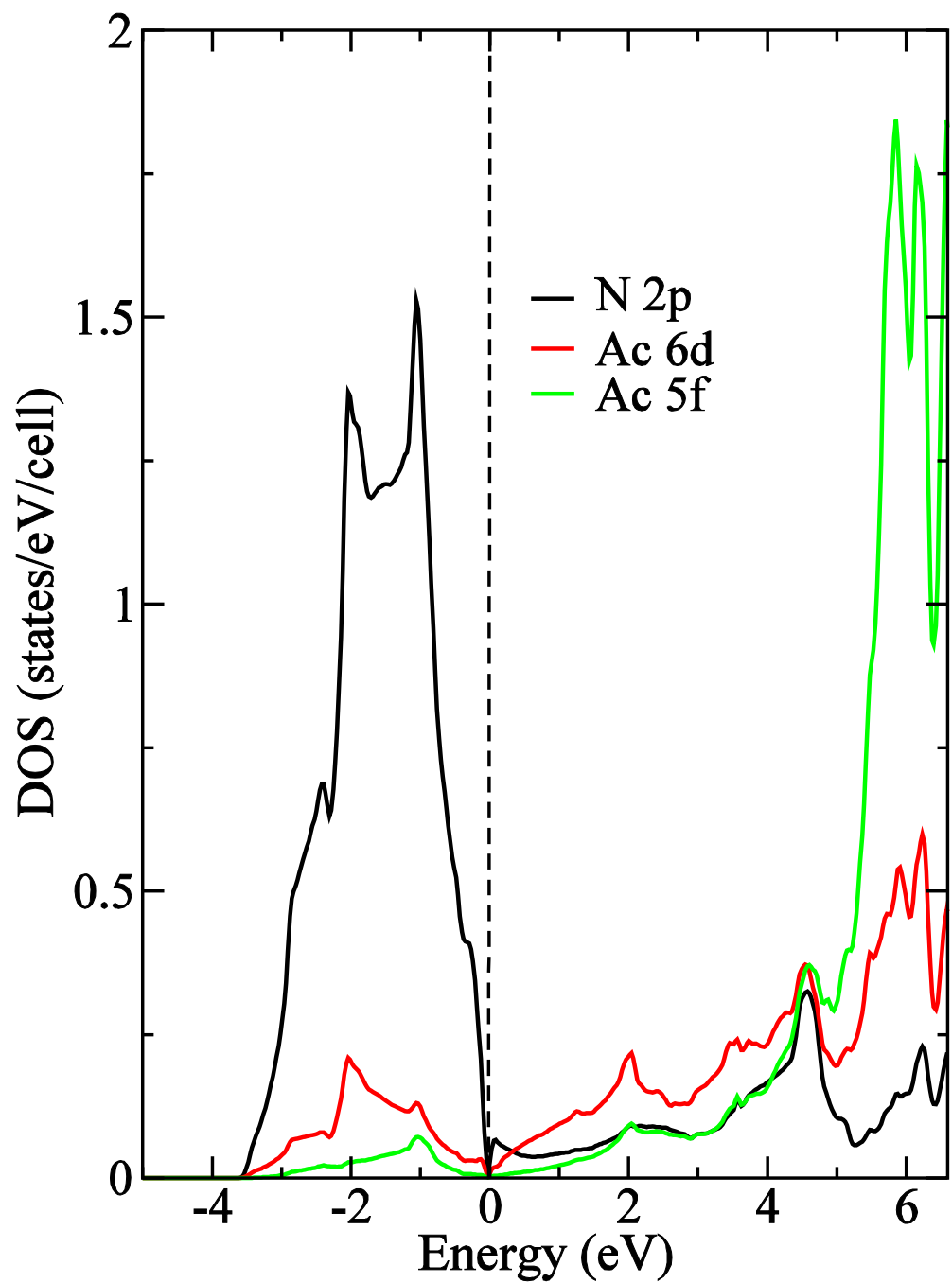

Figure 5 (Color Online): Partial DOS for AcN. Vertical line through E=0 is the Fermi level.

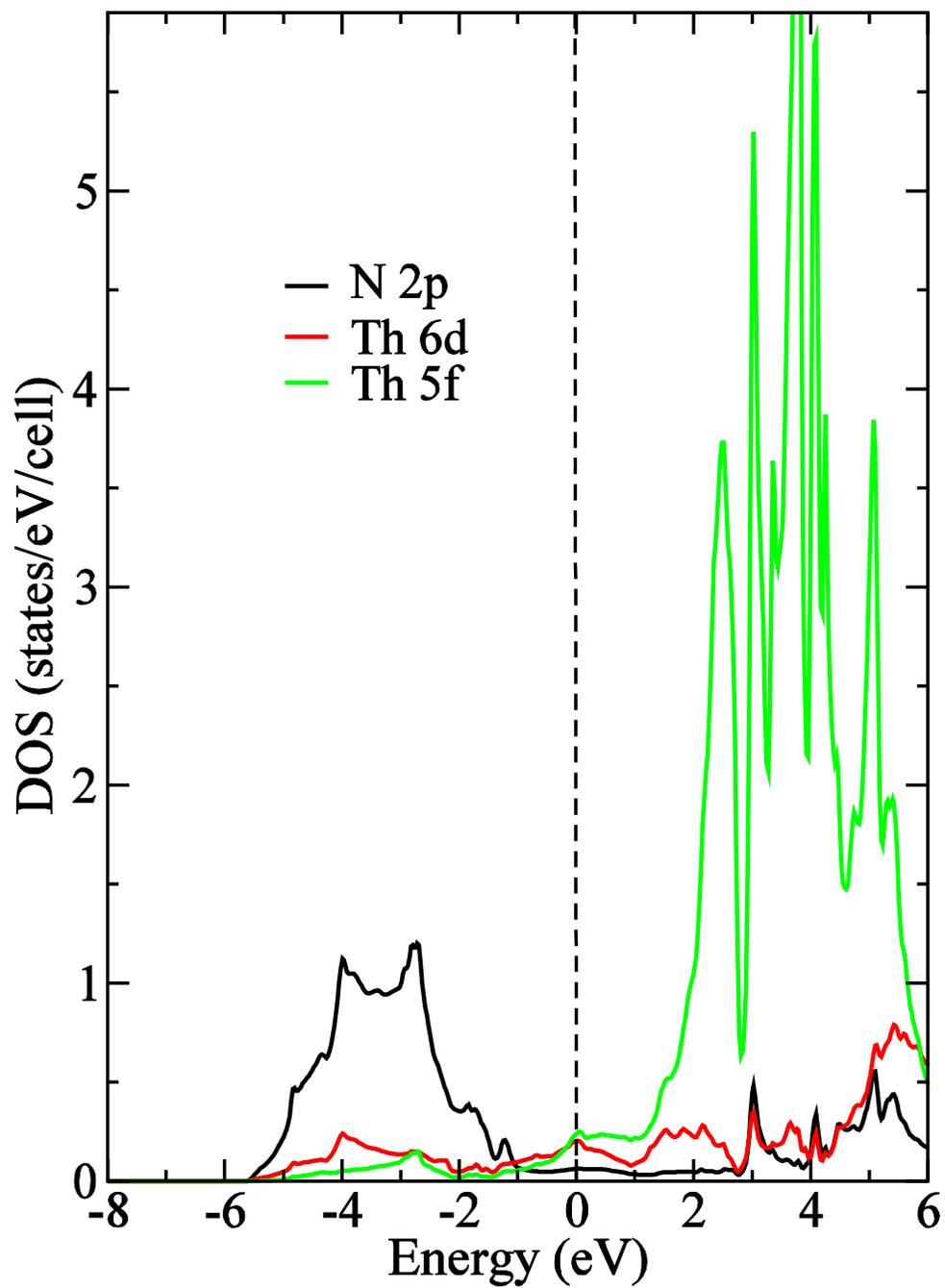

Figure 6 (Color Online) : Partial DOS for ThN. Vertical line through E=0 is the Fermi level.

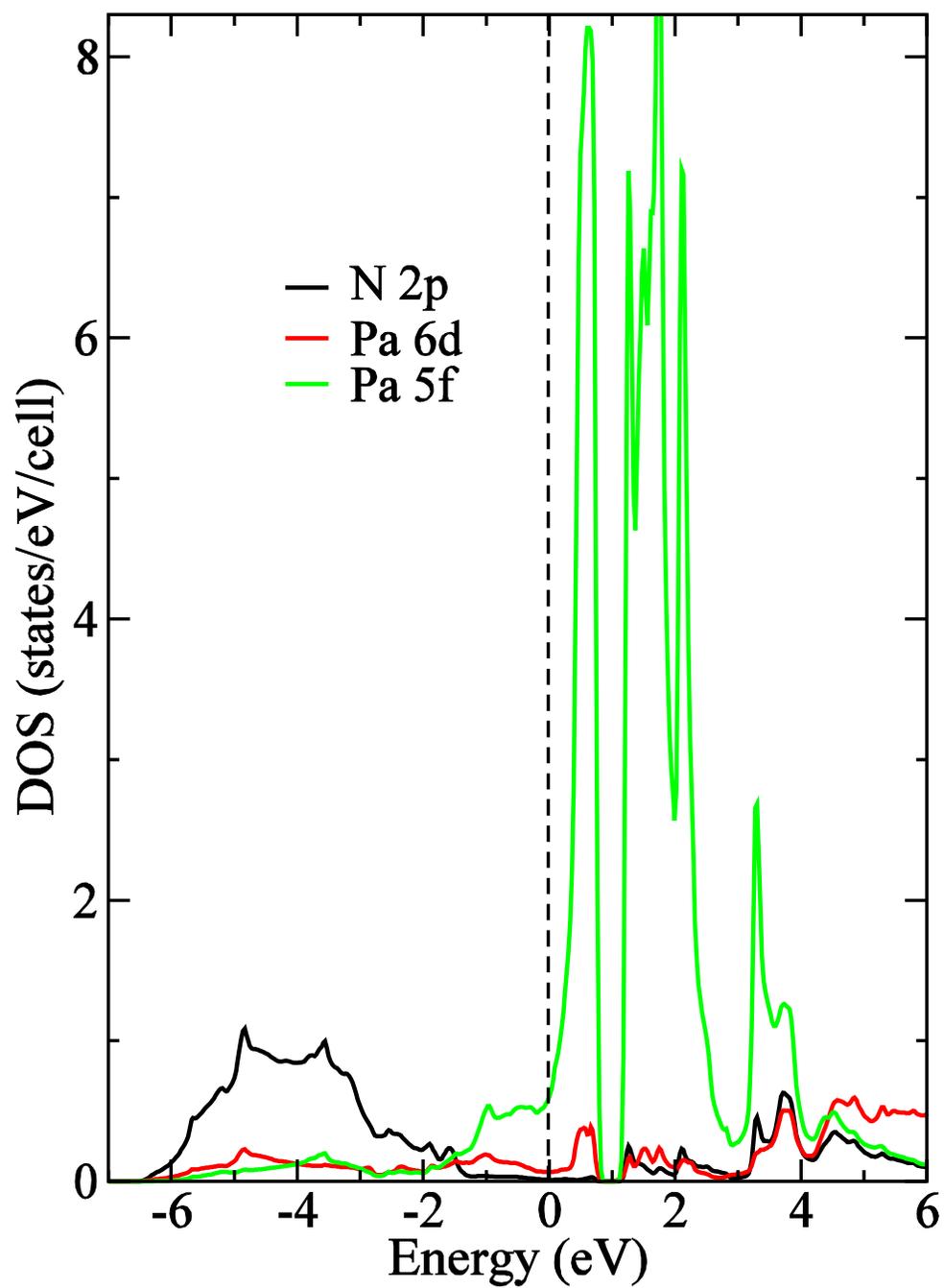

Figure 7 (Color Online): Partial density of states for PaN. Fermi level is the vertical line through E=0.

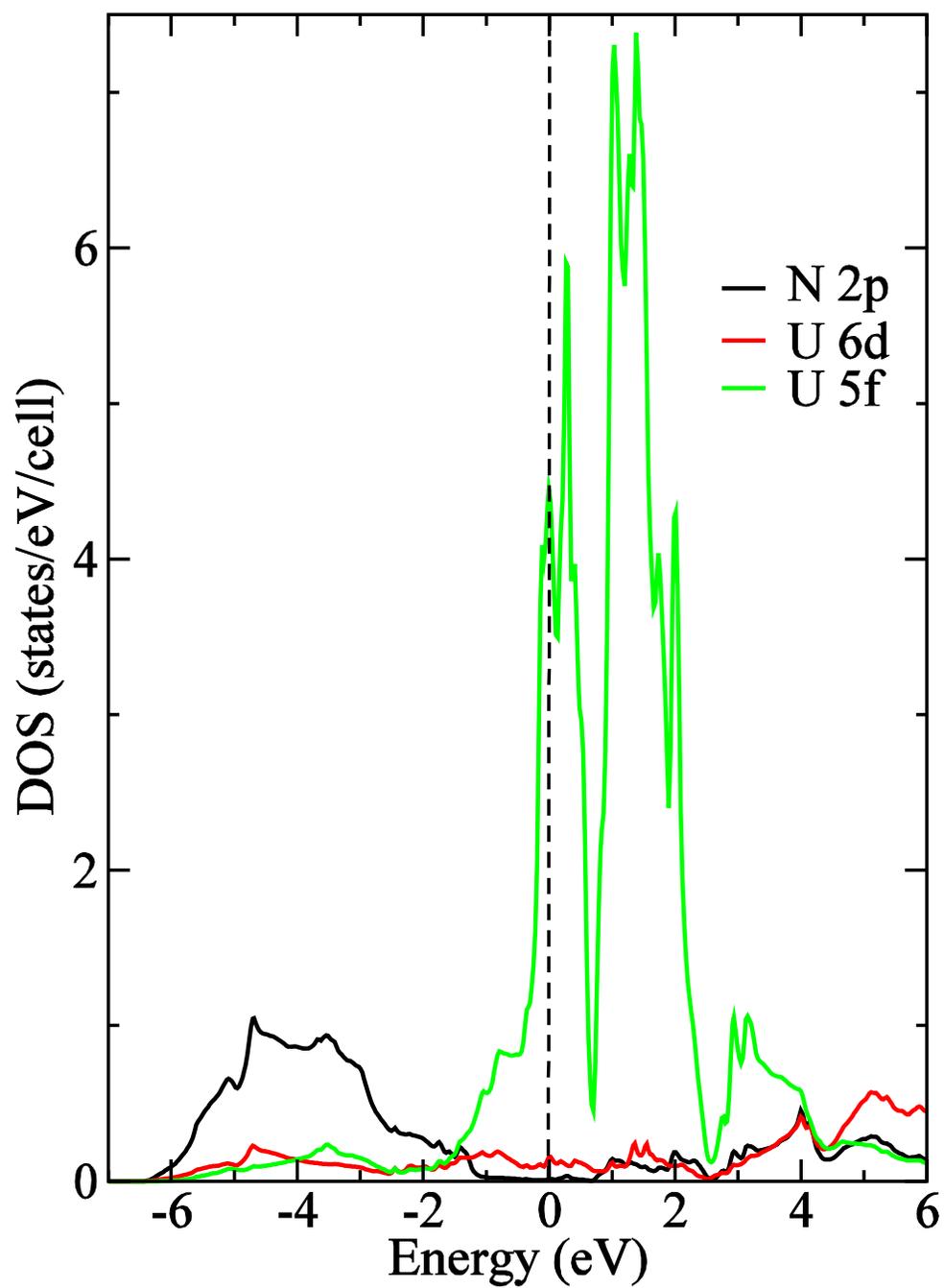

Figure 8 (Color Online): Partial DOS for UN. Vertical line through E=0 is the Fermi level.

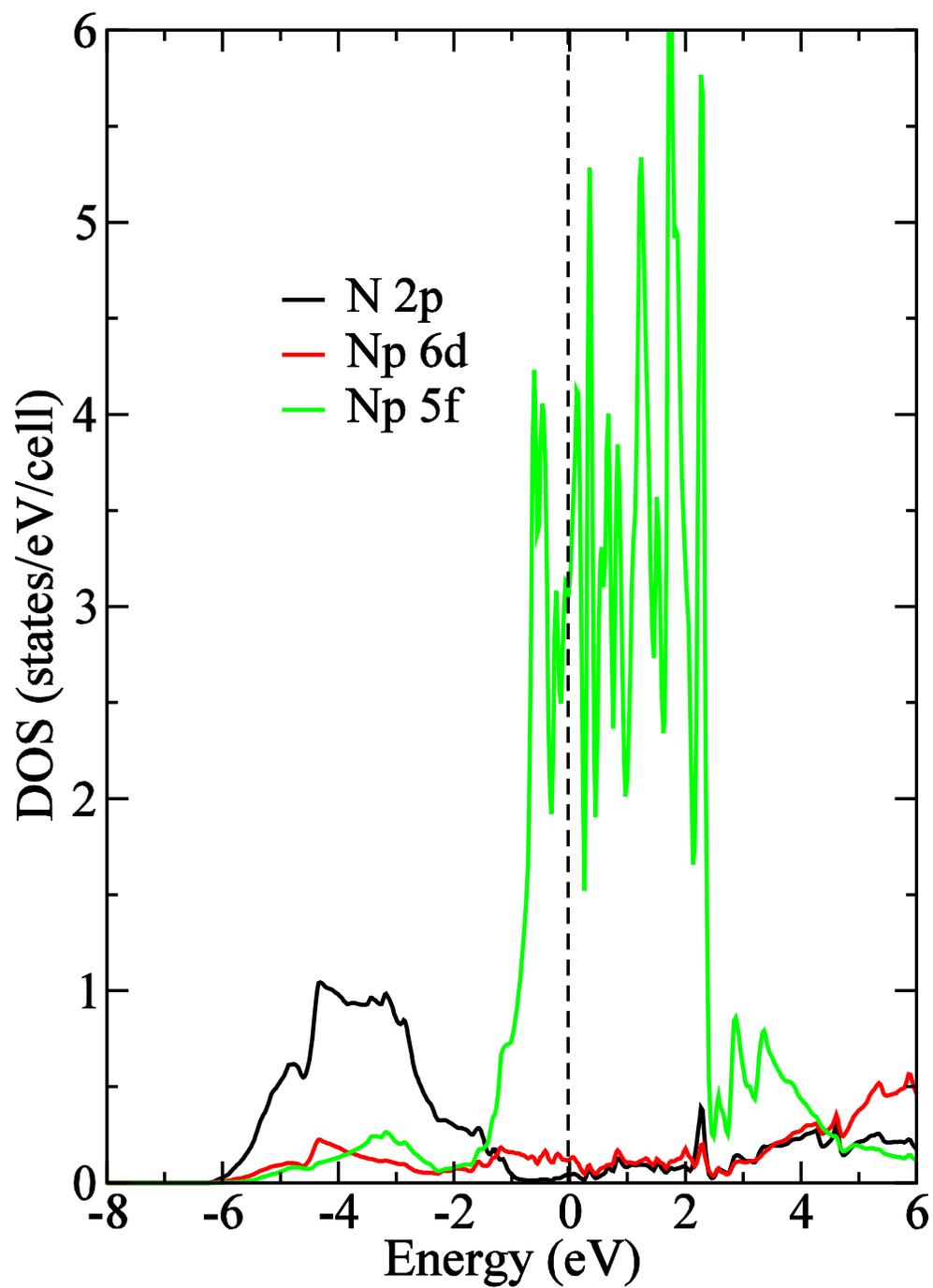

Figure 9 (Color Online): Partial density of states for NpN. Fermi level is the vertical line through E=0.

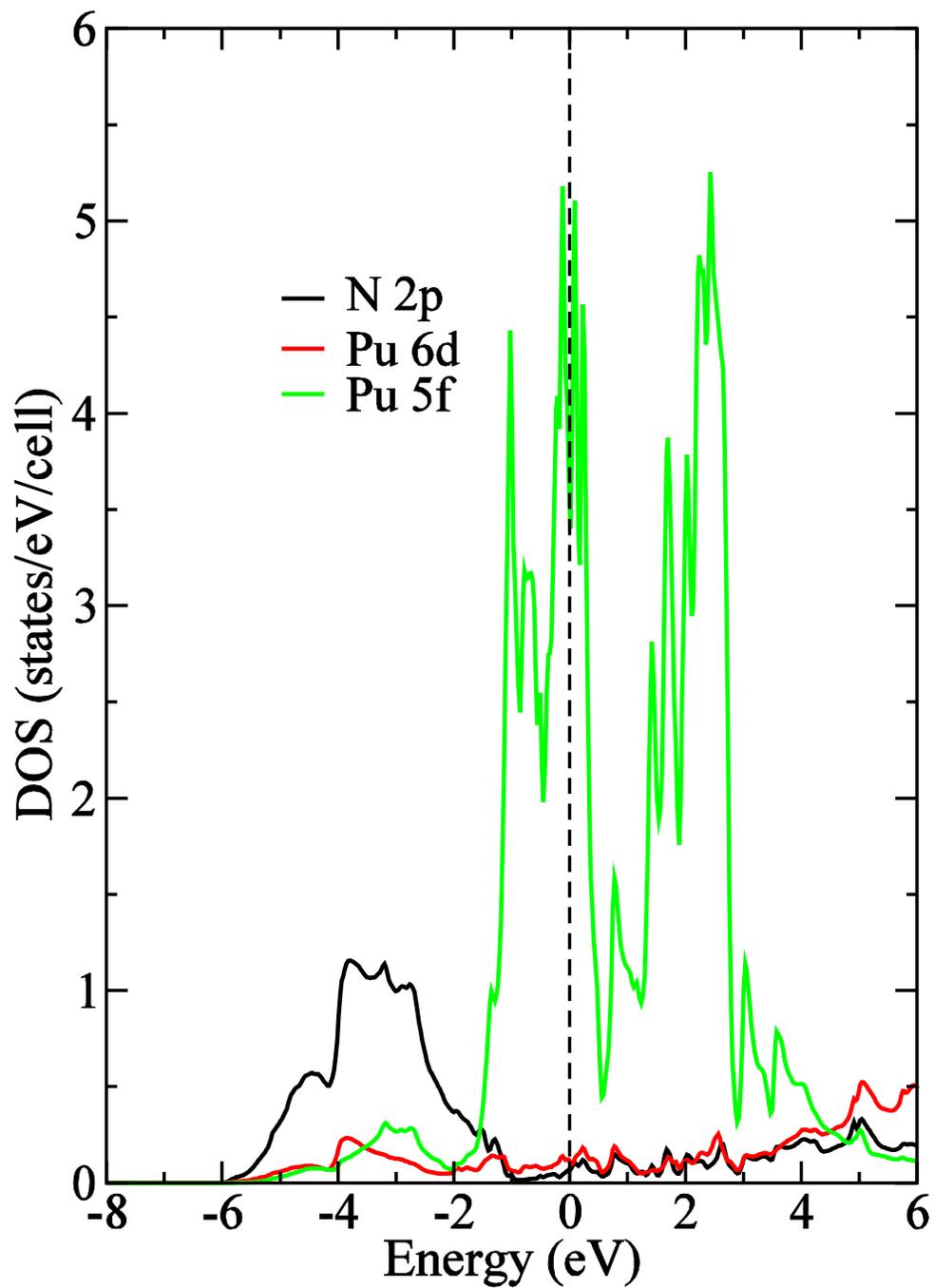

Figure 10 (Color Online): Partial density of states for PuN. Fermi level is the vertical line through E=0.

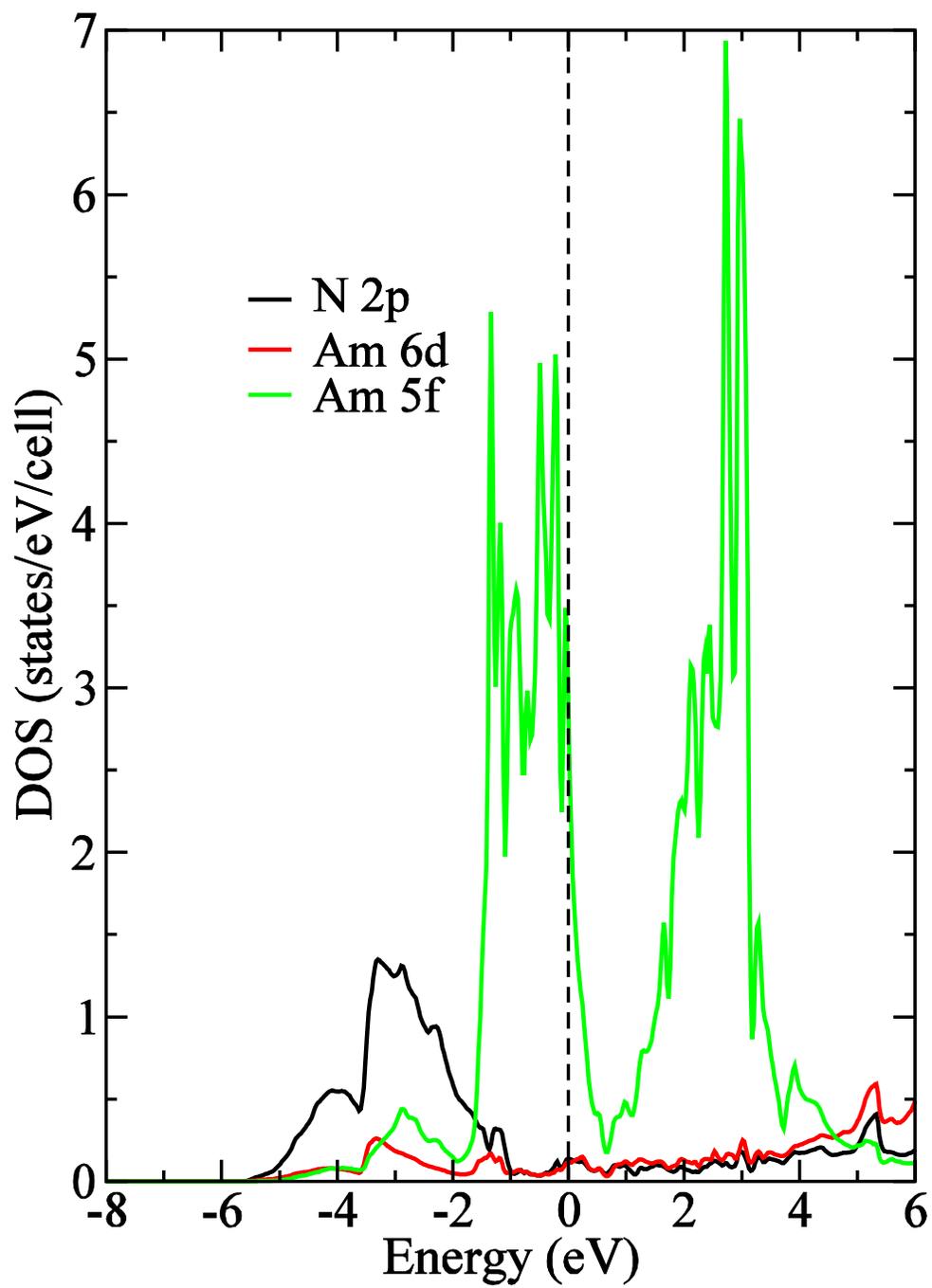

Figure 11 (Color Online): Partial density of states for AmN. Fermi level is the vertical line through E=0.

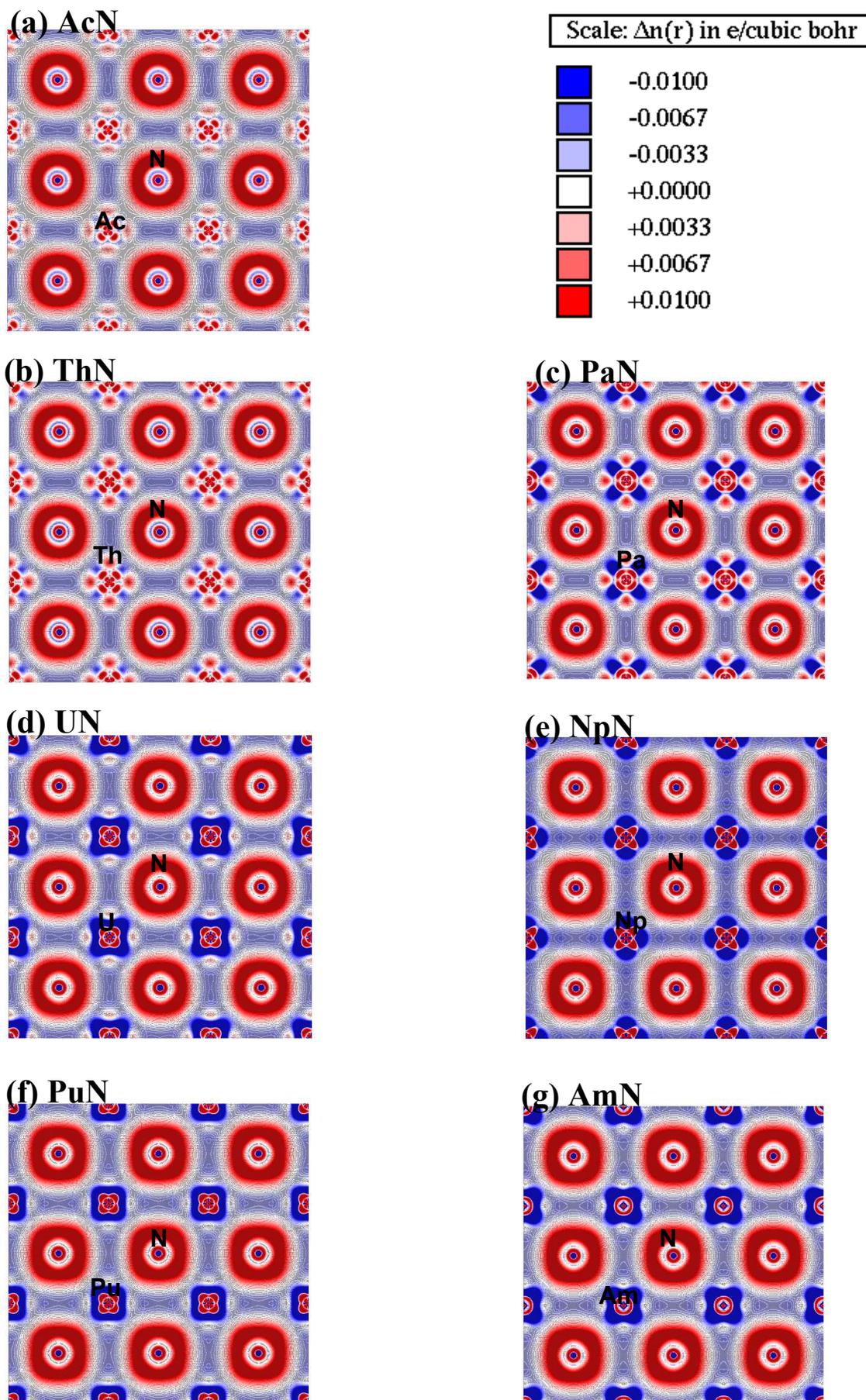

Fig 12 (Color Online): Difference electron density plots for (a) AcN, (b) ThN, (c) PaN, (d) UN, (e) NpN, (f) PuN, (g) AmN computed in the (001) plane. Atoms are labeled accordingly and the scale used is indicated at the top. Regions colored red represent charge gain and regions colored blue represent charge loss.